\newcommand\SF[4]{{{}^{#1}\hat\Psi^{#2}_{#3,#4}}}

\documentclass[12pt]{article}
\begin{document}

%%%%%%%%%%%%%%%%%%% title page %%%%%%%%%%%%%%%%%%%%%%%%%%%%%%%%%%%%
\begin{titlepage}
\nopagebreak
\begin{flushright}
May 2001\hfill
UT-940\\
hep-th/0105175
\end{flushright}

\renewcommand{\thefootnote}{\fnsymbol{footnote}}
\vfill
\begin{center}
{\Large BCFT and Sliver state}%\\

\vskip 20mm

Yutaka MATSUO\footnote{
{\tt E-mail address: matsuo@phys.s.u-tokyo.ac.jp}
}
\vskip 1cm
Department of Physics, University of Tokyo\\
Hongo 7-3-1, Bunkyo-ku\\
Tokyo 113-0033\\
Japan
\end{center}
\vfill

\begin{abstract}
We give a comment on the possible r\^ole of
the sliver state in the generic boundary conformal
field theory. We argue that for each Cardy state,
there exists at least one projector in the string field theory.
\end{abstract}

\vfill
\end{titlepage}

One of the long standing problem in the string theory is the
description of the geometry.  Generically we do not need
the space-time at all to define
the consistent string background.  What is necessary instead is
the modular invariant representation of the (super) Virasoro
algebra with the certain central extension.
If we need the geometry, we have to extract it in principle from the
algebraic data of the CFT, namely the set of the primary fields,
their OPE coefficients, the modular invariant combination of
the left and right movers and so on.

In the development of the noncommutative geometry
\cite{r-CDS, r-SW}, a hint
to this problem is given in the context of the noncommutative
soliton\cite{r-GMS}.  In this approach the description of the space-time
is replaced by the (in general noncommutative) $C^*$-algebra $\cal A$.
The noncommutative soliton is defined as the projection operator
of $\cal A$.  Immediately after the discovery, it is 
interpreted as the D-brane in the presence of the background $B$
field \cite{r-HKLM}.

In the noncommutative geometry, a canonical way to extract the geometry
from the algebraic data is through the study their $K$-group
\cite{r-Connes, r-WO} which are classified into
two types, $K_0(\cal A)$ and $K_1(\cal A)$.  The latter is
identified as the isomorphic class of the unitary operator in
$\mbox{Mat}_{N\times N}(\cal A)$ for large enough 
$N$.  On the other hand 
the former one is represented by the projection operator 
in $\mbox{Mat}_{N\times N}(\cal A)$
which fits naturally with the very definition 
of the noncommutative soliton.
In the commutative geometry, the D-brane charge is
classified by the topological $K$-group
\cite{r-Witten1, r-Horava}.  The noncommutative
soliton gives the representation of $K$-homology group of the
operator algebra and thus describes the $D$-brane
in the noncommutative situation\cite{r-Matsuo, r-HM}.

Since the current examples of the noncommutative geometry
is restricted to the rather simple spaces such as
the Moyal plane or the noncommutative torus, it is desirable
to extend such framework to the full string theory. 
In this language the operator algebra $\cal A$ should be
replaced by the full (boundary) CFT module.
This is, actually, the philosophy advocated by Witten
long ago \cite{r-Witten} (see also \cite{r-Witten2}) in his 
open string field theory.

Recently there is a remarkable progress toward this direction
\cite{r-RZ,r-KP, r-RSZ, r-RSZ1, r-RSZ2, r-GT}. It is based on the
conjecture that we may find the ghost string field which satisfies,
\begin{equation}
 Q\Psi_{gh}=\Psi_{gh}\star \Psi_{gh}
\end{equation}
and if one factories the full string field as
$\Psi=\Psi_{m}\otimes\Psi_{gh}$, the equation of
motion for the matter part becomes,
\begin{equation}\label{e-projector}
 \Psi_m\star \Psi_m = \Psi_m\,\,,
\end{equation}
namely it is the projector with respect to Witten's $\star$-product.
Furthermore there is an explicit representation of
the projector \cite{r-KP} which was identified with the 
sliver state found in \cite{r-RZ}.  It is conjectured that
it describes  the $D25$-brane. In \cite{r-RSZ2,r-GT} the
star product is reformulated as the matrix multiplication
(see also \cite{r-Witten}) and the sliver state is identified as the
rank 1 projector in this matrix algebra.

In the boundary conformal field theory description,
the D-brane is described by the boundary state (Cardy state).
It interesting to see to what extent the information of 
Cardy state is encoded in the solution to (\ref{e-projector}).

We start from the general (rational) conformal field theory
with the set of the primary fields $\phi_i$ labeled by the set 
$i\in\cal I$ (we use the notation of \cite{r-BPPZ}).
The information of the module over the primary field
is encoded in
Ishibashi state \cite{r-Ishibashi}$|i\rangle\!\rangle$
which satisfies,
\begin{equation}
 (L_n-\bar L_{-n})|i\rangle\!\rangle=0,
\qquad
 \langle\!\langle i| \tilde q^{\frac{1}{2}(L_0+\bar L_0 -\frac{c}{12})}
|j\rangle\!\rangle
= \delta_{ij}\chi_i(\tilde q)\,\,,
\end{equation}
where $\chi_i(\tilde q)$ is the character of the irreducible
representation associated with the primary field $\phi_i$
and $\tilde q=e^{2\pi i/\tau}$.  In order to give a well-defined 
Hilbert space sum in the open string sector, we need to take
the linear combination of the Ishibashi state (the Cardy state)
\cite{r-Cardy},
\begin{equation}\label{e-Cardy}
 |a\rangle =\sum_{j\in {\cal E}} \frac{\psi^j_a}{\sqrt{S_{1j}}}
 |j\rangle\!\rangle
\end{equation}
where $S_{ij}$ is the modular transformation matrix of the
character ($1$ represents the identity primary field),
$\psi^j_a$ is the coefficients which is determined by the
constraints in the following and $a\in {\cal V}$ is
the set of labels for the boundary state.
${\cal E}$ (exponent) is the subset of ${\cal I}$ 
which has the diagonal term in the modular invariant partition
function \cite{r-BPPZ}.
The inner product between the boundary state is given 
after the modular transformation,
\begin{equation}\label{e-openstringsector}
 \langle b | \tilde q^{\frac{1}{2}(L_0+\bar L_0-\frac{c}{12})}
|a\rangle =\sum_{i\in{\cal I}} \chi_i(q)n_{ia}^b,
\end{equation}
with $q=e^{2\pi i\tau}$ and
\begin{equation}
 n_{ia}^b=\sum_{j\in \cal E} \psi^j_a (\psi^j_b)^* 
\frac{S_{ij}}{S_{1j}}.
\end{equation}
In order that (\ref{e-openstringsector}) have 
well-defined interpretation as the trace over open string
Hilbert space the coefficients $n_{ia}^b$ must be
non-negative integers.  
The matrices $(n_i)_a^b= n_{ia}^b$ 
should satisfy the Verlinde fusion algebra \cite{r-Verlinde},
\begin{equation}
 n_i n_j = \sum_{k\in\cal I} N_{ij}^k n_k\,\,,
\qquad
N_{ij}^k=\sum_{\ell\in \cal I}\frac{S_{i\ell}
S_{j\ell}S_{k\ell}^*}{S_{1\ell}}\,\,.
\end{equation}
% One of the claims in \cite{r-BPPZ} is that
% given the integer matrix $n_i$ which satisfies this algebra,
% one may define the Cardy state.  Namely the coefficients
% $\psi_a^j$ can be identified with the common eigenvectors
% of $n_i$ 
% \begin{equation}
%  (n_i)_a^b \psi_b^j = (S_{ij}/S_{1j})\psi_a^i\,\,.
% \end{equation}

As a consequence, the boundary fields 
${}^b\Psi^a_{j,\beta}$ need to have
four labels,  $j\in \cal I$ which represents the chiral
operator, $a,b\in\cal V$ for the two boundaries and 
$\beta=1,\cdots,n_{ia}^b$ to specify the redundancy in
this channel.  When $j=1$, namely when the chiral
field is the identity operator, there is the natural
restriction $a=b$ and $n_{1a}^b=1$.

For each boundary fields, one may assign the corresponding
open string field as,
\begin{equation}
 {}^b\Psi^a_{j,\beta}\rightarrow {}^b{\hat\Psi}^a_{j,\beta}
\end{equation}
which is defined by inserting the field ${}^b\Psi^a_{j,\beta}$
at the origin of the upper half plane 
and perform the path integral on the half disk
\cite{r-Witten}.

As explicitly discussed in section 3 of \cite{r-RZ}, one may define
the star product between the string field thus defined without
the explicit knowledge of the oscillator representation.
Indeed the three string vertex operator can be completely specified
by the infinite set of the conformal Ward identities,
\begin{equation}
 \langle V_3 | \oint_{\cal C}v(z) T(z)dz=0,
\end{equation}
where ${\cal C}={\cal C}_1+{\cal C}_2+{\cal C}_3$  is
the contour around three punctures and $v(z)$ is the 
holomorphic vector fields in the interior of the vertex.

It implies that the star product between the string associated with
the boundary operator can be derived from the operator product
expansion for the boundary operator \cite{r-BPPZ},
\begin{equation}\label{e-prod}
 {}^b\Psi^c_{i,\alpha_1}(x_1) {}^c\Psi^a_{j,\alpha_2}(x_2)
 =  \sum_{p,\beta, t} {}^{(1)}F_{cp}\left[
\begin{array}{c c}
 i&j\\ b & a
\end{array}\right]^{\beta\ t}_{\alpha_1\ \alpha_2}
\frac{1}{x_{12}^{\Delta_i+\Delta_j-\Delta_p}}
 {}^b\Psi^a_{p,\beta}(x_2)+\cdots,
\end{equation}
where $\Delta_i$ is the dimension of the $i$th chiral field
and ${}^{(1)}F$ is the $3j$ symbol for the boundary field
fusion algebra.
This algebra should be modified into the star product
algebra for the string fields,
\begin{equation}\label{e-star}
 \SF{b}{c}{i}{\alpha_1}\star \SF{c}{a}{j}{\alpha_2}
= \sum_{p,\beta,t,\left\{n\right\}} {}^{(1)}\tilde{F}_{cp}\left[
\begin{array}{c c}
 i&j\\ b & a
\end{array}\right]^{\beta\ t}_{\alpha_1\ \alpha_2}(\left\{n\right\})
\SF{b}{a}{p}{\beta}(\left\{n\right\}),
\end{equation}
where the summation over $\left\{n\right\}$ stands for the 
string field associated with the descendants. ${}^{(1)}\tilde{F}$ is
the coefficients which should be derived from ${}^{(1)}F$ 
but their explicit forms will not be necessary
at the level of this paper.

To define the projector for the star product is reduced to
the problem of finding them in (\ref{e-star}),
which is rather hopeless unless we know the good handling
of the $3j$ symbols. At this point, we would like to indicate
that there is an easy but seemingly meaningful solution. 
Let us consider ${}^a\Psi^b_{1,\alpha}$ which is associated with 
the identity operator. In this case, we need to put
$a=b$ and the index $\alpha$ can be
omitted since $n_{1a}^a=1$ .  If we write ${\bf 1}^a\equiv {}^a\Psi^a_1$,
eq.(\ref{e-prod}) is written as,
\begin{equation}
 {\bf 1}^a(x_1){\bf 1}^b(x_2)= \delta_{ab}{\bf 1}^b(x_2)\,\,.
\end{equation}
While it is the projection at the level of the operator product,
it is quite nontrivial to find the projector at (\ref{e-star})
because of the inclusion of the descendants.
However, quite remarkably, this is exactly the situation considered
in section 6 of \cite{r-RZ}.  Namely the $n\rightarrow \infty$ limit
of the conformal mapping,
\begin{equation}\label{e-map}
 w=\left(\frac{1+iz}{1-iz}\right)^{2/n}
\end{equation}
gives a finite transformation in terms of the Virasoro operators,
\begin{equation}
U^{sliver}\equiv \exp\left(-\frac{1}{3}L_{-2}+\frac{1}{30}L_{-4}
-\frac{11}{1890}L_{-6}+\cdots\right)
\end{equation}
which defines the infinitely thin wedge (sliver)\footnote{
The identity operator ${\cal I}^a$ in the string field theory 
($n=1$ case in (\ref{e-map})) gives another solution to this problem
satisfying
${\cal I}^a\star {\cal I}^b=\delta_{ab}{\cal I}^a$.  At this point,
we have no strong reason that the sliver state is better suited
than  this one except for the arguments in \cite{r-RSZ1,r-RSZ2}. 
We would like to come back to this problem in our future study.}.
We apply this operator to the string field associated with 
${\bf 1}^a$ and denote the corresponding
field as $\Xi^a$.  The claim in \cite{r-RZ} can be generalized 
in our situation as
\begin{equation}\label{e-Xi}
 \Xi^a\star\Xi^b=\delta_{ab}\,\Xi^b\,\,.
\end{equation}

Our argument may be summarized as follows.  For each Cardy state
(D-brane) labeled by $a\in\cal V$,  there always exists identity
operator ${\bf 1}^a$ in the $a$-$a$ sector of the open string
and one can construct a projector $\Xi^a$ 
in the string field theory as the sliver state.
This result seems to give a possible partial 
answer to the general expectation that D-branes can be obtained as
the solution to the open string field theory.
It may also imply that the D-brane charge
is classified by the $K$-group
of the boundary conformal field theory.

At this point we would like to make a few comments
which seems to be relevant to the future study.
\begin{enumerate}
\item We suppose that the solution describes
the D-brane of the index $a\in\cal V$
since it does nothing but project the boundary to $a$.
If there is another solution, it defines the projective
module over the D-brane and thus describes the D-branes
which may appear after tachyon condensation \cite{r-Sen}
(see also \cite{r-hist} for the early attempts).
 \item From the Virasoro algebra view point, it is more
natural to consider the projector 
into the irreducible representation which appeared
in the different context in \cite{r-BPPZ},
\begin{equation}\label{e-proj2}
 \Pi_{i}\equiv\sum_{\left\{n\right\}}|i\left\{n
\right\}\rangle
\langle i\left\{n\right\}|\,\,,\quad
\Pi_i\Pi_j=\delta_{ij}\Pi_j
\end{equation}
where $|i\left\{n\right\}\rangle$ is the
normalized Virasoro descendant states from the highest weight state
$|i\rangle$ and the summation is taken over all the 
Virasoro module. This projector satisfies 
\begin{equation}
L_n \Pi_{i}=\Pi_{i} L_n\,,\quad\mbox{and}\quad
\mbox{Tr}(\Pi_i\,q^{L_0-c/24})=\chi_i(q)
\end{equation}
and it may be regarded as
the analogue of the Ishibashi state.
It is natural to guess
that there is a linear transformation 
between $\Pi_i$ and $\Xi^a$ (or possibly ${\cal I}^a$) 
which is similar to (\ref{e-Cardy})
but at this moment it is difficult
to find the explicit form.
The problem comes from the fact that string field we constructed
is the element of the Hilbert state and not 
the operators acting on it.  The split string formalism
\cite{r-Witten,r-RSZ2,r-GT,r-KO} seems to give an important hint to this issue.
% However, it depends critically on the oscillator representation
% and at this moment it is difficult for us to use 
% In any case, the projector (\ref{e-proj2}) depends on the
% index of the irreducible representation and we need to change it
% to that of the boundary field.
\item An interesting question is what is the trace
of the projector. To seek the analogy with the noncommutative
soliton \cite{r-HKLM}, it seems natural to impose the trace to 
be one to get a single D-brane.
While it is conjectured to be true for the sliver \cite{r-RSZ1,r-RSZ2},
the question is  whether this is the necessary constraint.
The difference between the usual noncommutative
soliton and the string field theory projector
is that we need only the zero mode in the former and
the full Virasoro module in the latter.

If the string algebra belongs to the
type I von Neumann factor (roughly speaking the matrix algebra)
it is quantized and takes the integer value.  We think that it
is not generally true.  In the noncommutative theory on the
quantum torus \cite{r-qtorus}, we encountered type II factor
and the trace of the projector 
takes the continuous value.  If it belongs to
type III, the trace becomes ill-defined.  This may be possible
since, for example,
the trace of the projector (\ref{e-proj2}) is infinite.
The subtle point is whether to include the conformal descendants
and is related to the definition of the ghost string field
\cite{r-RSZ1, r-RSZ2}.
\item More sophisticated use of the projectors to classify the 
representation of the Virasoro algebra was recently considered
in \cite{r-PZ}. They applied Ocneanu's paragroup 
\cite{r-Ocneanu,r-subfactors}
to BCFT. Ocneanu's method is to use the
fusion algebra to define an infinite sequence of von Neumann factors.
Accordingly the basic data of the paragroup is the fusion
algebra and the $6j$-symbols which satisfy the analogue of
Moore-Seiberg relations\cite{r-MS}. It will be quite interesting
to find a direct relation between the projectors in the string
field theory and generators of Ocneanu's double triangle algebra
\cite{r-PZ} if it may lead to give a classification of the projector
of the fusion algebra (\ref{e-prod}).
\end{enumerate}

\noindent{\sl Acknowledgement:}
The author is indebted to T. Eguchi for giving a critical 
comment on the possibility of the 
noncommutative soliton in the Virasoro module.

The author is supported in part by Grant-in-Aid (\#13640267)
and in part by Grant-in-Aid for Scientific Research
in a Priority Area ``Supersymmetry and Unified Theory of 
Elementary Particle'' (\#707) from the Ministry of Education,
Science, Sports and Culture.

\vskip 5mm

\noindent {\sl Note added:} After we have written this letter,
we found that the similar issue is discussed in \cite{r-RSZ3}.

\end{document}